# Improved Eigenvalue-based Spectrum Sensing via Sensor Signal Overlapping

Liping Du
School of Computer & Communication Engineering
University of Science & Technology Beijing,
Beijing, P.R. China

Mihir Laghate, Chun-Hao Liu, Danijela Cabric
Electrical Engineering Department,
University of California, Los Angeles
Los Angeles, CA 90095-1594, USA

*Abstract*—Eigenvalue-based detectors are considered as an important method of spectrum sensing since they do not require the information about the primary user (PU) signal. In this paper we propose a method to improve the performance of the eigenvalue-based detector. The proposed method introduces a new test statistic based on combinatorial matrix with components which are overlapping subgroups extracted from the array of received signals. As a result, its covariance matrix has a larger maximum eigenvalue and trace value than the one without overlapping. Simulation results show that our proposed method can further improve the detection performance of the optimal eigenvalue-based detector. The paper also shows the effect of different overlapping methods on the receiver operating characteristic curve.

## I. INTRODUCTION

Cognitive Radio (CR) provides more efficient communication by allowing secondary users (SUs) to utilize the licensed band which is allocate to the primary users (PUs). In cognitive radio networks, the secondary users have to detect the presence of PU accurately before they can utilize the channel to avoid interference with PUs.

Several spectrum sensing algorithms including the energy detection (ED), matched filter detection, and feature-based detection have been proposed and discussed in terms of their performance complexity tradeoffs to detect the primary transmitter [1], [2]. Recently, the methods based on random matrix theory (RMT), including the covariance-based detection [3] and eigenvalue-based detection [4] have been proposed. Eigenvalue-based detector can be further divided into two classes: detectors with known noise variance and detectors with unknown noise variance. The nearly optimal test statistic for the two classes is the largest eigenvalue test for known noise variance and the ratio of largest eigenvalue to trace of the covariance matrix for unknown noise variance, respectively [5]. The first one is also called Roy's largest root test (RLRT), proposed in [6], [7] and the latter is introduced in CR as a generalized likelihood ratio test (GLRT) [8]. There are also some other major eigenvalue-based detection techniques studied in the literature, i.e., maximum–minimum eigenvalue (MME) detection [4] and energy with minimum eigenvalue (EME) detection [9]. These spectrum sensing methods work well without the information about the channel, the primary signal or the noise variance.

In this paper, we propose a method to improve the detection performance of the conventional eigenvalue-based cooperative spectrum sensing schemes. In the conventional eigenvalue-based spectrum sensing scheme, it often adopts eigenvalues from the covariance matrix of received signals to build the test statistic. However, in our proposed method, a combinatorial matrix obtained from received signals is used to generate the covariance matrix and the corresponding eigenvalues. The combinatorial matrix has components which are overlapping subgroups extracted from the array of received signals. In the proposed method, the test statistic could be the same as the conventional eigenvalue-based spectrum sensing schemes. After overlapping, the covariance matrix of the combinatorial matrix has larger maximum eigenvalue and trace value than the non-overlapping covariance matrix. This results in higher signal-to-noise ratio (SNR) for the received signals. Moreover, higher SNR makes the probability density functions of test statistic under $H_0$ and $H_1$ further away from each other, where $H_0$ and $H_1$ are the hypothesis without and with PU signals, respectively. This is helpful to improve the detection performance. Simulation results show that our proposed improvement method can further improve the detection performance of the optimal eigenvalue-based detector. We also simulate the influence of the SNR and the overlapping number on the detection performance.

The rest of the paper is organized as follows. Section II introduces the system model and the overlapping method. In Section III, eigenvalue-based detection methods are presented. Section IV shows the simulation results. Finally we conclude the paper in Section V.

## II. SYSTEM MODEL

Assume there are one primary and $M$ secondary users/sensors in a cognitive radio network. The received data matrix can be expressed as

$$\mathbf{X} = \mathbf{A} \cdot \mathbf{S} + \mathbf{W} \qquad (1)$$

where the received data $\mathbf{X} = \{x_{mn}\}$ is an $M \times N$ matrix, $N$ is the number of samples. The $1 \times N$ vector $\mathbf{S} = \{s_n\}$, $n = 1,2,\ldots,N$ denotes transmitted signal from the primary users. The $M \times 1$ vector $\mathbf{A} = \{a_m\}$ represents the channels between

This work is supported by a National Science Foundation Project of P. R. China (No.61202079) and in part by the scholarship from China Scholarship Council (CSC) (No. 201306465007).

the primary users and $M$ sensors, $m = 1,2,...,M$. $\mathbf{W} = \{w_m\}$ is assumed to be $M \times N$ additive white Gaussian noise (AWGN). The noise of each sensor is zero mean and variance $\sigma_n^2$ and is uncorrelated with the received signal at each sensor.

As shown in Fig.1, $M$ sensors are divided into $p$ overlapping subgroups. Then each subgroup is an $(M - p + 1) \times N$ matrix, which can be expressed as

$$\mathbf{X_1} = \mathbf{A_1} \cdot \mathbf{S} + \mathbf{W_1}$$

$$\mathbf{X_2} = \mathbf{A_2} \cdot \mathbf{S} + \mathbf{W_2}$$

$$\vdots$$

$$\mathbf{X_p} = \mathbf{A_p} \cdot \mathbf{S} + \mathbf{W_p}$$

where

$$\mathbf{A_1} = \{a_{m_1}\}, \quad m_1 = \{1,2,...,M-p+1\},$$

$$\mathbf{A_2} = \{a_{m_2}\}, \quad m_2 = \{2,3,...,M-p+2\},$$

$$\vdots$$

$$\mathbf{A}_p = \{a_{m_p}\}, \quad m_p = \{p,...,M\},$$

$$\mathbf{W_1} = \{w_{m_1 n}\}, \mathbf{W_2} = \{w_{m_2 n}\}..., \mathbf{W_p} = \{w_{m_p n}\}.$$

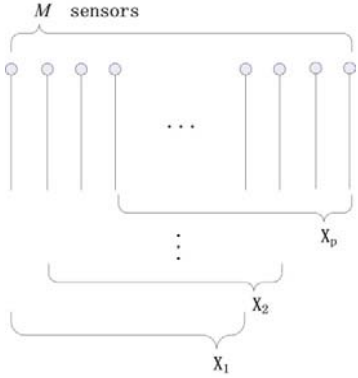

Fig. 1. Overlaying subgroups. $M$ sensors are devided into $p$ overlapping subgroups.

We aggregate the subgroup matrix and form a new combinatorial matrix

$$\mathbf{X}_c = [\mathbf{X_1 X_2} \cdots \mathbf{X}_p]. \qquad (2)$$

Our new data matrix $\mathbf{X}_c$ is a complex rectangular block Hankel matrix. The eigenvalues of such a matrix have not been studied in literature. Under hypothesis $H_0$, all elements of $\mathbf{X}_c$ are identical and independent distribution (i.i.d.) Gaussian distributed because only noise is received. Such matrices have been studied in literature under certain conditions described next. It is shown in [10] that the distribution of the eigenvalues of a real Hankel matrix with i.i.d. entries is symmetric, has unbounded support, and is not unimodal. The spectral radius of a Hankel matrix consisting of i.i.d. elements with positive mean is shown to converge to the mean in [11]. A semicircle law has been derived in [12] for real square block Hankel matrices whose blocks are square and have equal sizes.

The sample covariance matrix of $\mathbf{X}_c$ is defined as

$$\mathbf{R}' = \tfrac{1}{N} \mathbf{X}_c \mathbf{X}_c^\mathbf{H}. \qquad (3)$$

Define the sample covariance matrix of $\mathbf{X}$ as

$$\mathbf{R}_{M \times M} = \frac{1}{N}\mathbf{XX}^H = \begin{bmatrix} r_{11} & r_{12} & \cdots & r_{1M} \\ r_{21} & r_{22} & \cdots & r_{2M} \\ \vdots & \vdots & \ddots & \vdots \\ r_{M1} & r_{M2} & \cdots & r_{MM} \end{bmatrix}.$$

Then $\mathbf{R}'$ can be expressed as

$$\mathbf{R}'_{2M' \times 2M'} = \begin{bmatrix} r_{11} & r_{12} & \cdots & r_{1M'} & r_{12} & r_{13} & \cdots & r_{1M} \\ r_{21} & r_{22} & \cdots & r_{2M'} & r_{22} & r_{23} & \cdots & r_{2M} \\ \vdots & \vdots & \ddots & \vdots & \vdots & \vdots & \cdots & \vdots \\ r_{M'1} & r_{M'2} & \cdots & r_{M'M'} & r_{M'2} & r_{M'3} & \cdots & r_{M'M} \\ r_{21} & r_{22} & \cdots & r_{2M'} & r_{22} & r_{23} & \cdots & r_{2M} \\ r_{31} & r_{32} & \cdots & r_{3M'} & r_{32} & r_{33} & \cdots & r_{3M} \\ \vdots & \vdots & \cdots & \vdots & \vdots & \vdots & \ddots & \vdots \\ r_{M1} & r_{M2} & \cdots & r_{MM'} & r_{M2} & r_{M3} & \cdots & r_{MM} \end{bmatrix}.$$

Denote the ordered eigenvalues of $\mathbf{R}'$ as $\lambda_1 \geq \lambda_2 \geq \cdots \geq \lambda_{2M'}$ and $M' = M - p + 1$.

We use binary hypothesis test to decide whether primary users are absent or not. Let hypothesis $H_1$ represent primary users' signals presence while $H_0$ for absence. In the case of a single primary user, the hypothesis test can be expressed as

$$H_0: \mathbf{X} = \mathbf{W} \qquad (4)$$

$$H_1: \mathbf{X} = \mathbf{A} \cdot \mathbf{S} + \mathbf{W}. \qquad (5)$$

To make a decision on the status of primary users, we can compare the test statistic function $T$ with a predefined threshold $\gamma$. In this paper, we use two near-optimal test statistics: Roy's largest root test (RLRT) and generalized likelihood ratio test (GLRT). The detailed information of the two test statistic is shown in the following section. If $T > \gamma$ we decide $H_1$, otherwise $H_0$. Then the probability of detection and probability of false alarm can be expressed as

$$P_d = \Pr(T > \gamma(\alpha)|H_1) \qquad (6)$$

$$P_f = \Pr(T > \gamma(\alpha)|H_0) \qquad (7)$$

where $\gamma(\alpha)$ is the decision threshold such that $P_f = \alpha$.

## III. EIGENVALUE-BASED SPECTRUM SENSING

In this section, we use two nearly-optimal test statistics: Roy's largest root test (RLRT) and generalized likelihood ratio test (GLRT) to test the effect of our proposed method on the performance of eigenvalue-based spectrum sensing.

### A. RLRT

In the case where the channel and the noise variance are known, the most nearly-optimal approach to detect the presence of primary users is the well-known Neyman-Pearson (NP) procedure which consists of rejecting the null hypothesis when the observed likelihood ratio is larger than a certain threshold [13]. For the optimal detector in Neyman-Pearson sense, the likelihood ratio test (LRT) statistic is

$$LRT = \frac{p(\lambda_1,...,\lambda_{2M'} | H_1)}{p(\lambda_1,...,\lambda_{2M'} | H_0)}.$$

According to [14], the logarithm of the likelihood ratio is

$$\log \frac{p(\lambda_1,...,\lambda_{2M'} | H_1)}{p(\lambda_1,...,\lambda_{2M'} | H_0)} = \frac{n}{2}[\lambda_1 \frac{\sigma_s^2}{\sigma_s^2+1} - \log(1+\sigma_s^2)](1+o(1))$$

where $\sigma_s^2$ is the variance of the signal. With signal variance $\sigma_s^2$ and noise variance $\sigma_n^2$, distinguishing between $H_0$ and $H_1$ only depends on the largest eigenvalue $\lambda_1$. Hence, the likelihood ratio optimal detector can be considered as Roy's largest root test, which is defined as [5]

$$T_{RLRT} = \frac{\lambda_1}{\sigma_n^2}. \quad (8)$$

### B. GLRT

When $\sigma_n^2$ is unknown, the NP procedure does not apply to the composite hypothesis. A classical approach is to use the GLRT, which is shown as [15]

$$GLRT = \frac{\sup_{A,\sigma_n^2} p(X | H_1)}{\sup_{\sigma_n^2} p(X | H_0)}. \quad (9)$$

Thus the GLRT can be defined as

$$T_{GLRT} = \frac{\lambda_1}{\frac{1}{L}tr(R)} \quad (10)$$

where $L$ is the dimension of $\mathbf{R}$.

As for the proposed combinatorial matrix $X_c$, (10) can be rewritten as

$$T_{GLRT} = \frac{\lambda_1}{\frac{1}{2M'}tr(R')}. \quad (11)$$

## IV. SIMULATION RESULTS

We assume there is one primary user in the cognitive radio networks. The primary user's signal used in the simulation follows a Gaussian distribution with mean $\mu_s$ and variance $\sigma_s^2$, and the noise also follows a Gaussian distribution with mean $\mu_n$ and variance $\sigma_n^2$. The number of samples $N$ is 200. In the implementation of our algorithm, different numbers of sensors are used to evaluate the performances of detection. The following shows the effect of our proposed improvement method on the detection performance in three aspects, i.e., eigenvalues of covariance matrix, receiver operating characteristic (ROC) curves and the signal to noise ratio (SNR). Fig.2 is obtained with 50 Monte Carlo realizations and Fig.2-7 are run over 10000 realizations.

### A. Eigenvalues of covariance matrix

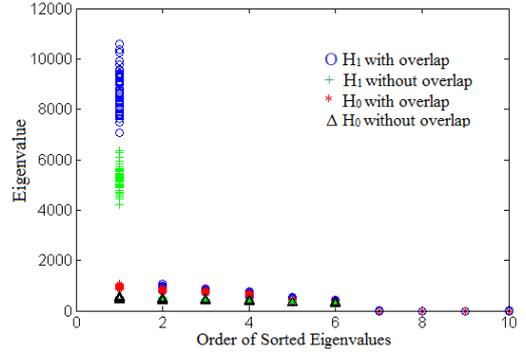

Fig. 2. The sorted eigenvalues of the sample covariance matrices with six receiver antennas. The blue circles and red asterisks are the eigenvalues of the proposed sample covariance matrix $\mathbf{R}'$ with overlapping under $H_1$ and $H_0$, respectively. The green crosses and black triangles represent the eigenvalues of $\mathbf{R}$ without overlapping under $H_1$ and $H_0$, respectively.

Fig.2 gives an example of the sorted eigenvalues of four different sample covariance matrix and SNR=3dB. We use six receiving sensors and $p=2$. The number of eigenvalues is 10. In Fig.2, it is clear that there is one dominant eigenvalue under $H_1$ due to one primary user's signal. Under $H_1$, the dominant eigenvalue of $\mathbf{R}'$ is generally larger than that of $\mathbf{R}$ while other eigenvalues of $\mathbf{R}'$ are almost the same as $\mathbf{R}$. Under $H_0$, the eigenvalues of $\mathbf{R}'$ are very close to that of $\mathbf{R}'$. The result shows that on average, eigenvalues of $\mathbf{R}'$ has larger gap between $H_1$ and $H_0$ than $\mathbf{R}$. If we use the eigenvalues of the covariance matrix as the test statistic, $\mathbf{R}'$ would have a better detection performance than $\mathbf{R}$ with a proper threshold.

In Fig. 2, it is shown that the covariance matrix of the combinatorial matrix has larger trace value than that of covariance matrix without overlapping. The larger trace results in larger SNR of received signal. Hence the two probability density functions of $H_0$ and $H_1$ will move further apart when a target is present, which is helpful to improve the detection performance of spectrum sensing.

*B. ROC*

The following results are simulated with 8 receiving sensors and the SNR is -13dB. In Fig.3, we show the ROC curves using the RLRT detector with overlapping number $p$ varying from 1 to 7 ($p=1$ means no overlapping). RLRT detector without overlapping has a slightly better detection performance than RLRT with $p=7$. The RLRT detectors with $p=2$ to 6 have better detection performance than $p=1$ and 7. The ROC curves with $p=2$ to 5 perform similar. According to the ROC curve, we can observe the performance follows the descending order of $p$ as 3>4>2>5>6>1>7. And the order of $p$ can be affected by the SNR.

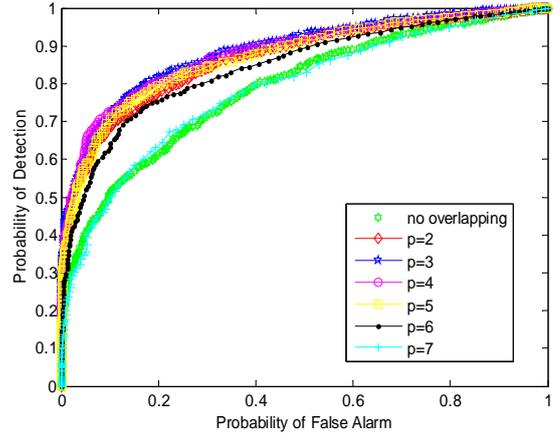

Fig. 4. ROC curves of GLRT detector with $p=1$ to 7 ($p=1$ means no overlapping). SNR=-13dB and $M$=8. The number of samples used for each sensor is 200.

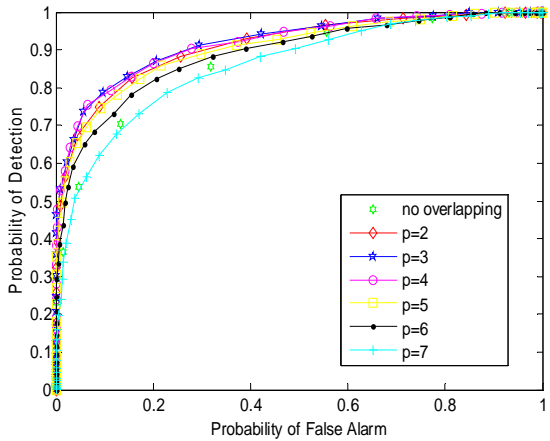

Fig. 3. ROC curves of RLRT detector with $p=1$ to 7 ($p=1$ means no overlapping). SNR=-13dB and $M$=8. The number of samples used for each sensor is 200.

We also simulate the detection performance of GLRT detector with different overlapping number $p$ under unknown noise variance. Due to one primary user existing in the cognitive radio network, there is one absolute high maximum eigenvalue while other eigenvalues are very low, which is shown in Fig. 2. Therefore, the detection performances of the GLRT detector have the same trend as RLRT detector, which is shown in Fig. 4. The GLRT detectors with $p=2$ to 6 have better detection performance than $p=1$ and 7. And the GLRT detector without overlapping has nearly the same detection performance as GLRT with $p=7$.

In Fig. 5 we compare the two optimal detectors with some other eigenvalue-based detector, such as, MME [4] and EME [9]. As for those detectors without overlapping, the detection performance follows the order of RLRT>GLRT>MME>EME. Under the same simulation condition, the combinatorial matrix with $p=2$ is calculated and applied to the RLRT and GLRT detector. The results show that our proposed improvement process can further improve the detection performance of the RLRT and GLRT detectors. In addition, we also apply the overlapping covariance matrix to other eigenvalued-based detectors, and their detection performances are all improved.

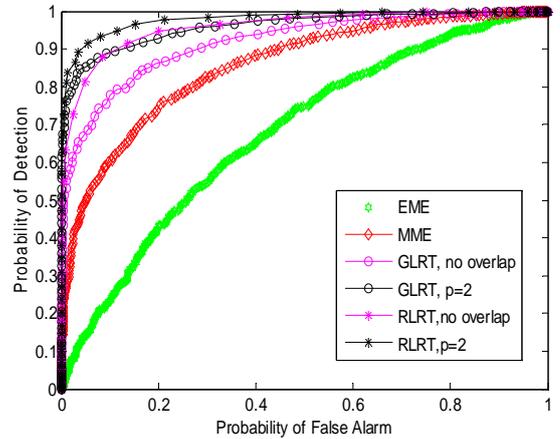

Fig. 5. ROC curves for different detection methods (SNR=-13dB,$M$=8). The proposed mehtod is compared with EME and MME. Under the same simulation condition, the combinatorial matrix with $p=2$ is calculated and applied to the RLRT and GLRT detector.

*C. SNR*

We fix the probability of false alarm at 0.1 and repeat the simulation 1000 times to get the SNR-$P_d$ curve with different overlapping number $p$. Fig.6 shows the impact of the SNR and the overlapping number on the probability of detection of the RLRT detector. Generally, the detection performance increases as the SNR increases. In the simulation, when SNR is lower than about -14dB, the RLRT with no overlapping has the lowest probability of detection than that with $p=2$ to 7. The RLRT with $p=7$ has the lowest detection performance than that of other $p$ values when SNR is larger than -14dB.

As for GLRT detector, the effect of SNR and overlapping number on the detection performance is shown in Fig.7. The trend of the curves is very similar to the RLRT detector. When SNR is larger than about -13dB, the $p=7$ performs worst. On the other hand, GLRT without overlapping performs worst when SNR is smaller than -13 dB. This means that low SNR problem can be improved by overlapping covariance matrix.

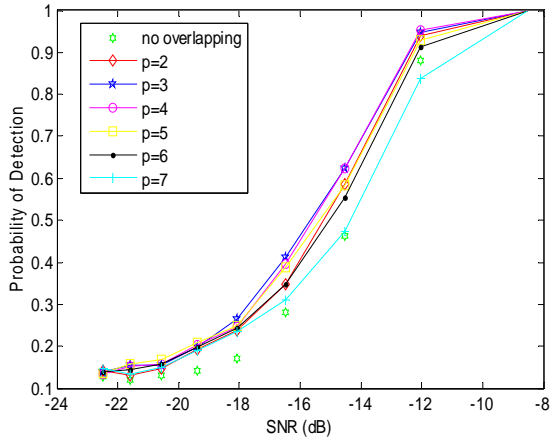

Fig. 6. Probability of detection versus SNR at $P_f$=0.1 using RLRT with $p$=1 to 7 ($p$=1 means no overlapping). The number of sensors $M$ is 8. The number of samples used for each sensor is 200.

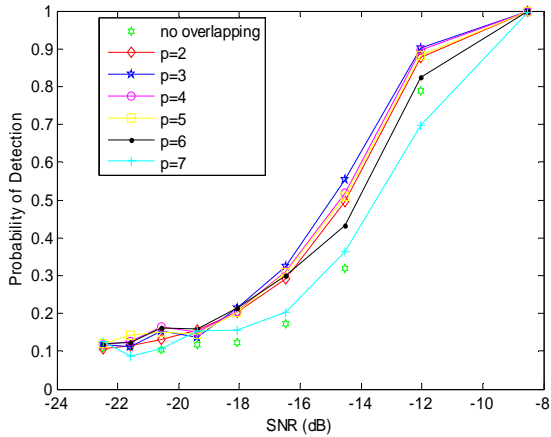

Fig. 7. Probability of detection versus SNR at $P_f$=0.1 ($M$=8) using GLRT with $p$=1 to 7 ($p$=1 means no overlapping). The number of sensors $M$ is 8. The number of samples used for each sensor is 200.

## V. CONCLUSIONS

Eigenvalue-based spectrum sensing methods have attracted much attention for their good detection performance. This kind of methods can detect primary users without requiring the information about the channel, the primary signal or the noise variance. In this paper, we propose an improvement method for eigenvalue-based detector. In our proposed method, the overlapping subgroups of received signal by multiple sensors are incorporated into a new combinatorial matrix. The new combinatorial matrix is used to derive the eigenvalue-based detector. We apply the improvement method to two optimal detectors, RLRT and GLRT. The simulation results show that our proposed method can improve the detection performance of the two detectors especially under low SNR scenario. We can also conclude that the overlapping number has different effects on the detection performance and we will theoretically analyze the effect of overlapping number in our future work.